\documentclass[10pt, a4paper, twocolumn]{article}

\usepackage[utf8]{inputenc}
\usepackage[T1]{fontenc}
\usepackage[english]{babel}

\usepackage[top=2cm, bottom=2cm, left=1.5cm, right=1.5cm]{geometry}

\usepackage{mathpazo} 
\usepackage{microtype} 

\usepackage{graphicx}
\usepackage{xcolor}
\usepackage{float}
\usepackage{subcaption} 
\usepackage[font=small, labelfont=bf, labelsep=period]{caption}
\usepackage{amsmath, amssymb, amsfonts}

\usepackage{authblk} 
\usepackage{titlesec} 
\usepackage{abstract} 

\usepackage{booktabs} 
\usepackage{multirow}
\usepackage{amsmath}

\usepackage[colorlinks=true, linkcolor=blue, citecolor=blue, urlcolor=blue]{hyperref}

\titleformat{\section}{\large\bfseries}{\thesection}{1em}{}
\titleformat{\subsection}{\normalsize\bfseries}{\thesubsection}{1em}{}

\newcommand{\corrauthor}[2]{\affil{\small Corresponding author: #1 (#2)}}

\title{\textbf{\Large Uncovering the Invisible in Routine Radiology: Cross-Modal Detection of MRI-Verified Soft-Tissue Pathologies from CBCT via Physics-Guided 3D Self-Supervised Learning}}

\author{Evgeny Alves Limarenko}

\affil{\small Moscow Institute of Physics and Technology, Dolgoprudny, Moscow Region, Russia}
\corrauthor{Evgeny Alves Limarenko}{e@alevlab.com}
\date{} 
\begin{document}

\twocolumn[
  \begin{@twocolumnfalse}
    \maketitle
    \begin{abstract}
      Routine volumetric X-ray modalities such as cone-beam computed tomography (CBCT) are optimized for high-contrast mineralized tissues, leaving soft-tissue abnormalities buried beneath photon noise and below the threshold of human visual perception. Consequently, detecting non-calcified pathologies---ranging from occult localized inflammation to soft-tissue neoplasms---traditionally requires secondary referral to magnetic resonance imaging (MRI). In this study, we demonstrate that visually imperceptible (\textit{sub-visual}) radiographic signatures of soft-tissue pathology are physically preserved within routine CBCT scans and can be extracted for opportunistic screening using a targeted, physics-guided deep learning pipeline, even under extreme paired-data scarcity. Using the temporomandibular joint (TMJ) as a clinical model system ($N = 212$ paired CBCT--MRI joints acquired within $\le 30$~days), we address the severe overfitting of high-capacity 3D networks ($86.28 \times 10^6$ parameters) through a two-stage framework. First, we employ a strictly frozen full-grid 3D masked autoencoder (BertsWin) pre-trained on 6,550 unlabeled CBCT volumes via the GradientConductor (GCond) optimizer and an anatomy-aware objective (PhysLoss) that explicitly redirects gradient flow into the {$[-300, 300]$}~HU soft-tissue attenuation window. Second, automated 3D localization of the mandibular condylar zenith via a hybrid Attention U-Net head drives a deterministic 3D cropping mechanism that eliminates 81\% of background osseous variance ($14^3 \to 8^3$ token grid). Evaluated via a patient-isolated $3 \times 10$-fold stratified cross-validation protocol with 3D affine training augmentations, a statistically robust sub-visual signal for MRI-verified synovial effusion ($\text{ROC AUC} = 0.793 \pm 0.036$, Youden's $J = 0.459$) emerges exclusively from representations pre-trained with the coupled PhysLoss and GCond regime. Crucially, this regime is the only pre-training configuration capable of capturing true 3D anatomical chirality ($p < 0.001$) and high intra-patient structural consistency ($0.791$) over inter-patient noise; conversely, identical architectures pre-trained with standard $L_2$ loss (which relies on isotropic texture statistics) or unmasked variance decomposition (MVC, where patch statistics remain dominated by high-contrast bone) collapse to random guessing on downstream effusion detection ($\text{ROC AUC} = 0.465\text{--}0.513$, $J \approx 0.07$). These findings establish a generalizable paradigm for repurposing routine, low-contrast radiographic scans into opportunistic screening tools for occult soft-tissue disease.
      
      \vspace{0.5cm}
      \vspace{0.5cm}
    \end{abstract}
  \end{@twocolumnfalse}
]

\section{INTRODUCTION}

In modern diagnostic radiology, high-throughput volumetric X-ray modalities---most notably cone-beam computed tomography (CBCT)---serve as the frontline standard for evaluating mineralized anatomical structures \cite{Li2022, Osama2025}. However, the physical principles of cone-beam image formation impose a severe diagnostic blind spot: wide-cone scatter radiation, truncated field-of-view artifacts, and low contrast-to-noise ratios render soft-tissue structures visually homogeneous and diagnostically uninformative to the human observer \cite{Liu2023}. As a result, non-calcified abnormalities---whether occult inflammatory processes, fluid effusions, or early soft-tissue lesions incidentally captured within the field of view---remain undetected during routine radiographic procedures. Definitive soft-tissue evaluation strictly requires magnetic resonance imaging (MRI) \cite{Jeon2021, Manek2025}, whose deployment as a population-wide screening tool is precluded by high operational costs, limited scanner accessibility, prolonged acquisition protocols, and patient contraindications \cite{Chalkoo2015}.

This clinical dichotomy is acutely exemplified in the diagnosis of temporomandibular joint (TMJ) disorders \cite{Jeon2021, Mehta2025}. While CBCT is routinely acquired in dental and maxillofacial workflows and achieves superior diagnostic accuracy and sensitivity over MRI in detecting osseous alterations such as cortical erosions and osteophytes (87\% vs. 68\% sensitivity, respectively) \cite{Jadeja2024}, it provides no direct visual delineation of the articular disc or synovial cavity \cite{Osama2025}. Clinicians relying on CBCT are forced to infer soft-tissue status indirectly from surrogate geometric measurements of the joint space width. While these linear measurements are highly reproducible in themselves (ICC > 0.90) \cite{Gorucu2019}, relying on them as indirect predictors for articular disc displacement suffers from limited diagnostic validity. Specifically, surrogate geometric evaluation exhibits high sensitivity and NPV, making it a viable preliminary screen, but severely lacks specificity (often below 50\%), resulting in a disproportionately high false-positive rate and low predictive power \cite{Osama2025, Fu2026}. Yet, from a physical standpoint, X-ray attenuation differentials within the soft-tissue Hounsfield Unit (HU) spectrum ($[-300, 300]$~HU) are not absent in CBCT; rather, they exist as subtle, visually imperceptible (\textit{sub-visual}) statistical variations buried beneath stochastic photon noise and overshadowed by high-magnitude osseous boundaries \cite{Dayarathna2024}. Analogous to how deep neural networks have uncovered occult systemic biomarkers from routine retinal photographs \cite{Poplin2018}, we hypothesize that a domain-tailored 3D computational pipeline can isolate these sub-visual soft-tissue signatures directly from routine CBCT scans, enabling opportunistic cross-modal screening of pathologies traditionally verifiable only via MRI.

Extracting these sub-visual signatures to build a robust predictive pipeline requires overcoming three fundamental methodological barriers:

First, training cross-modal 3D predictors is bottlenecked by the extreme scarcity of synchronously acquired paired datasets \cite{Zhou2021, Ma2024}. Because soft-tissue states such as synovial inflammation fluctuate over time, pairing routine CBCT scans with ground-truth MRI reports requires a tight temporal inclusion window ($\le 30$~days between examinations, compared to the 90-day intervals common in clinical practice). Under such rigorous filtering, verified clinical cohorts rarely exceed a few hundred volumes. In this extreme low-data regime, end-to-end supervised training of high-capacity 3D convolutional or transformer networks inevitably degenerates into memorizing stochastic background noise and patient-specific bone contours \cite{Zhou2021, Lee2023}.

Second, while self-supervised learning (SSL) via Masked Autoencoders (MAE) \cite{He2022} offers a powerful paradigm for learning anatomical priors from large unlabeled archives \cite{Huang2023}, its direct application to 3D medical volumes is structurally and mathematically bottlenecked. Standard sparse 3D MAEs discard 75\% of volumetric patches during encoding, severing continuous 3D anatomical neighborhoods. This topological discontinuity is a recognized fundamental challenge, forcing recent frameworks to mitigate the loss of spatial coherence through computationally complex workarounds, such as integrating sparse convolutions \cite{Wald2025} or employing dual-branch global-local feature aggregation modules \cite{Zhuang2025}. More critically, conventional MAE pre-training relies on voxel-wise mean squared error (L2 loss) \cite{Terven2025, Sara2019}. Across a 3000 HU dynamic range where soft tissues occupy a narrow fraction of the intensity spectrum, conventional L2 optimization disproportionately penalizes high-amplitude errors at high-density cortical bone interfaces. As recently demonstrated, MSE-based optimization acts as a low-pass filter for fine details, overwhelmingly preserving high-amplitude dominant edges (macro-geometry) while heavily blurring subtle high-frequency textural features \cite{Madan2025}. Consequently, encoders trained with standard L2 loss and applied to CBCT become hyper-focused on macroscopic osseous geometry, failing to resolve the critical, low-contrast textural gradients of intra-articular soft tissues necessary for opportunistic screening \cite{Yeung2022}. Consequently, $L_2$-trained encoders fail to learn oriented 3D morphological geometry, relying instead on isotropic local texture statistics \cite{LimarenkoBertsWin}.

Third, even if a self-supervised encoder successfully captures sub-visual soft-tissue features, feeding a complete peri-articular volume into a downstream classifier on a small paired cohort results in an unfavorable signal-to-noise ratio. The high spatial variance of adjacent cranial and mandibular bone structures overwhelms the subtle intra-articular signal unless the classification head is deterministically restricted to the relevant local anatomical region \cite{Isensee2021}.

In our companion methodological study, we introduced \textbf{BertsWin} \cite{LimarenkoBertsWin}, a full-grid hybrid 3D masked autoencoder that resolves topological sparsity by retaining all masked and visible tokens within a linear-complexity shifted-window transformer \cite{Liu2021}, coupled with the \textbf{GCond} optimizer \cite{Limarenko2026} and an anatomy-aware variance-decomposed objective (\textbf{PhysLoss}) based on Theil's decomposition \cite{Theil1961}. In that work, we showed that combining PhysLoss with GCond is the only pre-training regime that acquires statistically significant sensitivity to 3D anatomical chirality (distinguishing true contralateral left--right joints from synthetic reflections, $p < 0.001$) while maximizing intra-patient structural similarity ($0.791$) relative to inter-patient noise \cite{LimarenkoBertsWin}. In the present paper, we demonstrate that these exact structural properties are the prerequisite for extracting visually occult, MRI-verified soft-tissue pathologies from routine CBCT scans on a restricted cohort of 212 joints. The primary contributions of this work are threefold:
\begin{enumerate}
    \item \textbf{Discovery of Sub-Visual Soft-Tissue Signatures in CBCT:} We provide the first quantitative proof that MRI-verified synovial effusion can be detected directly from routine CBCT volumes with statistically and clinically significant accuracy ($\text{ROC AUC} = 0.793 \pm 0.036$, surpassing the Hosmer--Lemeshow clinical discrimination threshold of $0.70$ \cite{Hosmer2013}). Crucially, we prove via $3 \times 10$-fold cross-validation that this diagnostic signal is captured \textit{exclusively} by the chirality-aware PhysLoss + GCond encoder, whereas standard $L_2$ and unmasked multi-component variance (MVC) encoders collapse to random guessing ($\text{ROC AUC} = 0.465\text{--}0.513$).
    \item \textbf{Overfitting Mitigation via Frozen Probing and Deterministic 3D Token Cropping:} We show that any parametric adaptation of the $86.28 \times 10^6$-parameter backbone (including partial fine-tuning and Low-Rank Adaptation) induces overfitting in the low-data regime. By combining a strictly frozen encoder with an automated Attention U-Net localization module ($\text{ASSD} = 0.537$~mm) that crops an $8^3$ token sub-volume around the condylar zenith, we eliminate 81\% of background osseous variance and isolate the intra-articular diagnostic signal.
    \item \textbf{Delineation of Direct Soft-Tissue vs.\ Secondary Osseous Detection Limits:} By comparing synovial effusion against articular disc displacement, we establish the physical boundaries of cross-modal inference at $N = 212$ joints, demonstrating why volumetric fluid accumulation is directly detectable in the soft-tissue latent space while thin fibrocartilage displacement requires steering the classifier toward secondary cortical bone remodeling ($\text{ROC AUC} = 0.679$, $\text{Sensitivity} = 0.840$).
\end{enumerate}

\section{MATERIALS AND METHODS}

\subsection{Clinical Cohorts and Multi-Tiered Ground Truth Curation}

This retrospective study was approved by the Institutional Ethics Committee of the Skolkovo Institute of Science and Technology (Protocol dated April 17, 2026), and written informed consent was obtained from all participants prior to data collection. 
All volumetric DICOM files and clinical text records were de-identified prior to analysis in compliance with data protection regulations. 

\paragraph{Unlabeled SSL Pre-Training Archive.} 
Self-supervised representation learning was performed on the single-center clinical CBCT archive previously detailed in our companion study \cite{LimarenkoBertsWin}. Briefly, the dataset comprised 3,853 maxillofacial CBCT scans acquired at an isotropic resolution of $0.2$~mm, from which standardized $400 \times 400 \times 400$ voxel volumes ($80^3$~mm) centered on the TMJ were automatically extracted. Strict patient-level partitioning yielded 6,550 training joints and 1,156 validation joints to prevent data leakage. Voxel intensities were clipped to the $[-1011, 1992]$~HU interval and Z-normalized using statistics derived exclusively from the training split. Additionally, an external verification array of 2,199,444 head-and-neck CT slices from public repositories \cite{Flanders2020, Lin2023, Welch2024} was utilized strictly to evaluate gradient field stability across heterogeneous inputs.

\paragraph{Paired Multimodal Validation Cohort (CBCT + MRI).} 
To evaluate cross-modal detection of soft-tissue pathologies, we curated a clinical cohort of 99 patients (212 TMJs, including 14 longitudinal follow-up scans) who underwent both CBCT and MRI examinations. 
In routine clinical practice, the diagnostic interval between CBCT and MRI frequently reaches 90~days; however, to minimize label noise arising from natural disease progression or symptom resolution between scans, we enforced a strict inclusion threshold requiring an inter-scan interval of $\le 30$~days. 
To ensure that downstream classifiers did not overfit to the acquisition artifacts of a single device, the validation cohort incorporated CBCT volumes acquired across six different scanner models with isotropic voxel spacings ranging from $0.2$~mm to $0.4$~mm. 

\paragraph{Cross-Modal Label Extraction and Diagnostic Arbitration.} 
To avoid the subjectivity and annotation overhead of manual voxel-level delineation on low-contrast volumes, joint-level binary labels were obtained from the corresponding clinical MRI reports via automated extraction from unstructured clinical text.
Two primary diagnostic targets were formulated: (1) presence of \textbf{synovial effusion} (intra-articular fluid accumulation/inflammatory edema; positive-to-negative ratio $1:3$) and (2) \textbf{articular disc displacement} (positive-to-negative ratio $3:1$).
A third potential target, osseous destruction, exhibited a $1:4$ class imbalance, leaving only 4--5 positive instances per validation fold during 10-fold cross-validation; preliminary runs confirmed that this extreme sparsity caused systematic classifier collapse toward majority-class prediction, warranting its exclusion from the quantitative protocol. 
To guarantee label fidelity when parsing unstructured MRI reports, a three-tiered verification protocol was implemented: primary independent annotation by clinical residents, parallel automated diagnostic extraction using a large language model (Gemini Pro, Google, Mountain View, CA, USA), and final arbitration of all discrepancies---alongside blind quality-control sampling of concordant cases---by a senior gnathologist with over 25 years of clinical experience. 

\subsection{Physics-Guided 3D Self-Supervised Pre-Training}

To extract robust 3D features without supervision, we utilize our previously introduced \textbf{BertsWin} framework \cite{LimarenkoBertsWin}. 
Rather than repeating the exhaustive architectural benchmarks from our companion study, we summarize the core methodological design choices that directly govern downstream cross-modal sensitivity:

\paragraph{Full-Grid Hybrid Encoding.} 
Standard sparse 3D masked autoencoders (such as MONAI MAE-ViT \cite{Cardoso2022, He2022}) discard 75\% of volumetric patches during encoding, severing continuous 3D anatomical neighborhoods. This topological discontinuity is a recognized fundamental bottleneck in volumetric imaging, forcing recent state-of-the-art frameworks to mitigate the loss of spatial coherence through computationally complex workarounds, such as integrating sparse convolutions \cite{Wald2025} or employing dual-branch global-local feature aggregation modules \cite{Zhuang2025}. 
In contrast, BertsWin partitions a $224^3$-voxel input volume into non-overlapping $16^3$-voxel patches ($3.2^3$~mm in physical space), compresses the visible 25\% subset via a four-stage 3D convolutional stem (stride~2), and scatters the resulting embeddings back into a complete $14 \times 14 \times 14$ token grid ($2,744$ tokens, channel dimension $C = 768$) populated with learnable \texttt{[MASK]} tokens and harmonic positional encodings. 
This dense 3D token grid is processed by a 12-layer single-scale Swin Transformer \cite{Liu2021} utilizing $7 \times 7 \times 7$ shifted local windows (12 attention heads) without hierarchical downsampling. 
By restricting self-attention to local 3D neighborhoods, the encoder maintains linear $O(N)$ scaling while enabling iterative spatial signal relaxation across masked voids, preserving high-frequency anatomical interfaces prior to lightweight convolutional decoding. 

\paragraph{Multi-Component Variance (MVC) and Anatomy-Aware PhysLoss.} 
When applied to high-dynamic-range CT/CBCT volumes ($\sim 3000$~HU), conventional voxel-wise $L_2$ (MSE) loss prioritizes high-amplitude intensity errors at dense cortical bone boundaries, causing optimization of low-contrast soft-tissue structures to stagnate \cite{Sara2019, Terven2025}. 
Following Theil's decomposition \cite{Theil1961}, patch-level MSE between predicted ($X$) and target ($Y$) volumes can be algebraically partitioned into brightness (mean $\mu$), contrast (standard deviation $\sigma$), and structural (Pearson correlation $\rho$) components: 
\begin{equation}
\begin{split}
  \text{MSE}(X,Y) &= (\mu_X - \mu_Y)^2 + (\sigma_X - \sigma_Y)^2 \\
  &\quad + 2\sigma_X \sigma_Y \big(1 - \rho(X,Y)\big)
\end{split}
\end{equation}
To prevent brightness error from dominating gradient flow, we first formulate a weighted additive Multi-Component Variance objective: 
\begin{equation}
L_{\text{MVC}} = w_{\mu} L_{\text{Br}} + w_{\sigma} L_{\text{Cntr}} + w_{\rho} L_{\text{Str}}
\end{equation} 
where the weighting coefficients $w_{\mu} = 0.3$, $w_{\sigma} = 0.2$, and $w_{\rho} = 0.5$ were established via a dedicated ablation study on head-and-neck CT data \cite{LimarenkoBertsWin} to enforce concurrent convergence of structural correlation alongside radiometric intensity. 

However, because soft tissues occupy only a narrow fraction of the CBCT intensity histogram, unmasked patch-level statistics remain heavily biased toward osseous structures. 
To explicitly redirect encoder capacity toward intra-articular soft tissues, we introduce \textbf{PhysLoss}, which evaluates $L_{\text{MVC}}$ on $8^3$ local sub-patches across three dynamically masked anatomical domains: 
\begin{equation}
\begin{split}
  L_{\mathrm{phy}} &= \lambda_{\mathrm{global}}L_{\mathrm{MVC}}(\Omega) + \lambda_{\mathrm{soft}}L_{\mathrm{MVC}}(M_{\mathrm{soft}}) \\
  &\quad + \lambda_{\mathrm{surf}}L_{\mathrm{MVC}}(M_{\mathrm{surf}})
\end{split}
\end{equation}
Because an exhaustive grid search over regional weights is computationally prohibitive in 3D SSL pre-training, the coefficients $\lambda$ were assigned based on physical signal-equalization principles and TMJ anatomical priors: 
\begin{itemize}
    \item \textbf{Soft-Tissue ROI Mask ($M_{\text{soft}}$, $\lambda_{\text{soft}} = 0.5$):} Isolates voxels within the $[-300, 300]$~HU physical attenuation window (mapped to the corresponding Z-normalized intensity interval) representing intra-articular fluids and non-mineralized tissues.
    Because this regime suffers from the lowest contrast-to-noise ratio in CBCT, it is assigned the dominant weight ($0.5$) to amplify weak soft-tissue gradients and prevent feature collapse. 
    \item \textbf{Osseous-to-Soft-Tissue Interface Mask ($M_{\text{surf}}$, $\lambda_{\text{surf}} = 0.2$):} Generated via asymmetric morphological operations on the bone segmentation mask (inner erosion with kernel $k=2$ and outer dilation with $k=4$). 
    This shell captures both the thin cortical lamina and the immediate external transition zone into the joint cavity---a boundary critical for delineating articular surfaces and attempting to resolve the adjacent fibrocartilaginous disc interface, while decoupling surface topology from dense internal trabecular bone. 
    \item \textbf{Global Volumetric Regularizer ($\Omega$, $\lambda_{\text{global}} = 0.3$):} Computed across the complete unmasked patch domain to act as a spatial regularizer that preserves macro-structural continuity and background coherence. 
\end{itemize}

\paragraph{GCond Optimization and Latent Verification.} 
All self-supervised models were pre-trained with an effective batch size of $N = 192$ using our GCond optimizer \cite{Limarenko2026}. 
GCond integrates the sign-based momentum update of LION \cite{Chen2023, Rong2025} with the layer-wise adaptive trust-ratio scaling of LARS \cite{Do2024} and Adam-style first-moment bias correction:
\begin{equation}
    \hat{m}_t = \frac{m_t}{1 - \beta_1^t}
\end{equation}
By omitting second-moment variance tracking, GCond reduces the memory footprint of the optimizer state by 50\% compared to AdamW, while bias-corrected trust ratios stabilize sign updates across hybrid CNN--Swin layers. 
As demonstrated in our companion study \cite{LimarenkoBertsWin}, combining sign-based updates with layer-wise trust-ratio scaling enables the optimizer to force through transient gradient singularities where variance-accumulating optimizers (such as AdamW) stagnate, substantially accelerating convergence while driving the network into a deeper local optimum across both global and tissue-specific reconstruction metrics.
Prior to clinical deployment, the semantic integrity of the pre-trained encoders was audited on the 1,156-joint validation cohort via mean-centered cosine similarity probing, confirming that BertsWin (PhysLoss + GCond) is the only configuration that achieves both high intra-patient contralateral consistency ($0.791$) over inter-patient noise and statistically significant sensitivity to 3D anatomical chirality ($p < 0.001$, paired Wilcoxon signed-rank test comparing true contralateral joints against synthetically mirrored reflections) \cite{LimarenkoBertsWin}. 
\subsection{Condylar Zenith Localization and Signal-Focusing 3D Token Cropping}

\begin{figure}[ht]
    \centering
    \includegraphics[width=\columnwidth]{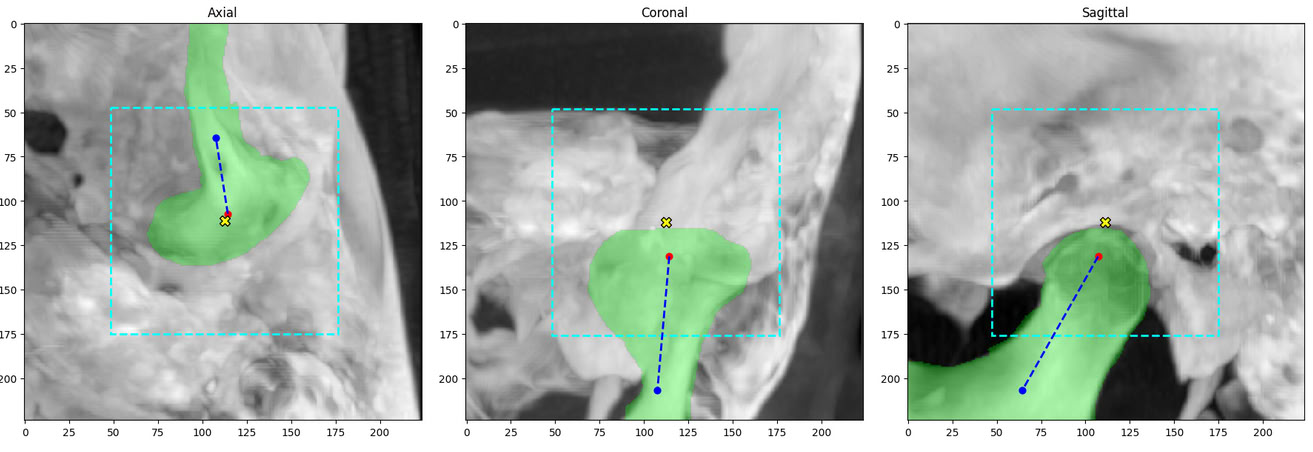}
    \caption{Automated condylar zenith localization and deterministic 3D cropping across orthogonal CBCT projections (axial, coronal, and sagittal). The green overlay depicts the predicted mandibular segmentation mask. Blue and red dots indicate the segmental centers of mass defining the mandibular orientation vector (dark-blue dashed line), which guides the identification of the superior condylar zenith (yellow cross). The cyan dashed bounding box delineates the localized $8 \times 8 \times 8$ token sub-volume ($128^3$~voxels) centered at the zenith.}
    \label{fig:crop_localization}
\end{figure}

When training a classifier on a restricted cohort ($N = 212$ joints), feeding the complete $14 \times 14 \times 14$ encoder feature map ($224^3$~voxels) proved ineffective in preliminary experiments: the subtle sub-visual signal originating from the intra-articular space was overwhelmed by the massive spatial variance of surrounding cranial and mandibular bone structures, yielding statistically inconclusive predictions. 
To boost the signal-to-noise ratio without manual intervention, we implemented an automated localization and deterministic 3D cropping pipeline (Fig.~\ref{fig:crop_localization}). 

To obtain reliable osseous boundaries for landmark extraction, we utilized a segmentation head on top of the strictly frozen BertsWin encoder, previously validated across the 1,156-joint cohort ($3 \times 5$-fold cross-validation) against expert-verified reference masks generated via DentalSegmentator \cite{Dot2024, Isensee2021}. 
As previously validated in our companion study \cite{LimarenkoBertsWin}, even with a lightweight two-layer non-linear decoder, the frozen BertsWin (PhysLoss + GCond) backbone reconstructs the mandibular condyle with $\text{DSC} = 0.829 \pm 0.003$ and an Average Symmetric Surface Distance ($\text{ASSD}$) of $0.744 \pm 0.125$~mm ($\text{DSC} = 0.927 \pm 0.002$, $\text{ASSD} = 0.317 \pm 0.061$~mm on the glenoid fossa). 
Building on these established anatomical representations, integrating skip connections from the frozen 3D CNN stem via a hybrid Attention U-Net head further refines condylar boundary precision in the current pipeline to $\text{DSC} = 0.947 \pm 0.007$ and $\text{ASSD} = 0.537 \pm 0.211$~mm ($\text{HD95} = 1.261 \pm 0.420$~mm).
At the native $0.2$~mm voxel spacing, a mean surface error of less than 3--4 voxels approaches the physical partial-volume limit of CBCT, ensuring highly reproducible landmark identification. 

From the predicted 3D binary mask of the mandibular condyle, our algorithm computes the segmental centers of mass to establish the primary mandibular orientation vector, analytically localizing the coordinates of the superior anatomical zenith of the condylar head. %
Using this zenith as a geometric anchor, we crop a localized $8 \times 8 \times 8$ token sub-grid (corresponding to a $128^3$-voxel physical sub-volume) from the full $14 \times 14 \times 14$ latent tensor. 
Rather than serving as a computational optimization, this spatial restriction discards 81.3\% of irrelevant macro-structural background ($2,744 \to 512$ tokens), forcing the downstream classification head to focus exclusively on the $\sim 19\%$ peri-articular volume housing the synovial cavity and articular interface. 

Crucially, the resulting zenith coordinates are shared across all evaluated backbones, ensuring that downstream comparisons isolate encoder representation quality under identical spatial cropping.

\subsection{Downstream Classification Heads and Cross-Validation Protocol}

Because the parametric capacity of the 3D encoder ($86.28 \times 10^6$ parameters) exceeds the paired clinical sample size ($N = 212$) by five orders of magnitude, unfreezing the backbone risks immediate memorization of individual osseous anatomy. 
While we systematically ablated partial fine-tuning (unfreezing the CNN stem and final transformer block), Low-Rank Adaptation (LoRA with ranks $r \in \{4, 8, 16\}, \alpha = 16$ and $r = 32, \alpha = 64$), and full end-to-end fine-tuning, our primary diagnostic pipeline freezes all encoder weights ($W_{\text{enc}} = \text{const}$) and restricts learning to lightweight, heavily regularized projection heads. 

\paragraph{Synovial Effusion Classifier.} 
To detect inflammatory effusion directly from the cropped $8 \times 8 \times 8 \times 768$ latent feature volume, we employ a non-linear gated spatial attention projector. The module compresses the 768 input channels to 64 via a $1 \times 1 \times 1$ convolution (with Group Normalization, GELU, and 3D spatial dropout $p = 0.2$), aggregates local $3 \times 3 \times 3$ neighborhood context via a residual depthwise-separable convolutional block, and computes gated spatial attention weights ($\tanh \odot \sigma$ gating followed by spatial softmax across the $8^3 = 512$ tokens) before a two-layer classification head ($64 \to 32 \to 1$ with Layer Normalization and dropout $p = 0.4$).

\paragraph{Articular Disc Displacement Classifier with Osseous Gradient Prior.} 
Unlike synovial effusion, which alters volumetric fluid attenuation across the joint capsule, the fibrocartilaginous articular disc exhibits X-ray attenuation nearly identical to adjacent musculature and ligaments. Consequently, direct classification of disc displacement from soft-tissue latent features alone failed to achieve statistical significance at $N = 212$, overfitting to background noise. To overcome this physical limitation, we implemented a dual-stream projection head that fuses the cropped $8^3 \times 768$ latent tokens (compressed to 32 channels via a $1 \times 1 \times 1$ convolution) with an explicit geometric stream extracted from the aligned $128^3$-voxel raw CBCT sub-volume. Specifically, the vertical finite-difference gradient $|\nabla_z I|$ and Z-normalized voxel intensities are downsampled to $16^3$ via area interpolation, concatenated with a 3-channel normalized 3D coordinate grid ($Z, Y, X \in [-1, 1]$), and mapped via a strided $3 \times 3 \times 3$ convolution ($\text{stride} = 2$) to a $32 \times 8 \times 8 \times 8$ tensor. This deterministic vertical-edge prior shifts classifier attention toward high-contrast cortical bone surfaces, where the concatenated 64-channel volume undergoes $1 \times 1 \times 1$ fusion, global max pooling, and linear classification (dropout $p = 0.3$).

\paragraph{Regularization and Statistical Validation.} 
To prevent overfitting within the projection heads, we combined Group and Layer Normalization, 3D spatial and dense dropout, label smoothing ($\alpha = 0.1$), and inverse-frequency weighted sampling within each training fold to compensate for the $1:3$ and $3:1$ class imbalances.
Generalization performance was evaluated using a patient-isolated $3 \times 10$-fold cross-validation protocol (30 total evaluation runs). 
The three repeats correspond to independent stratified 10-fold patient partitions (`StratifiedKFold` seeds), with training batches undergoing dynamic 3D affine augmentations (rotations, scaling, and translational shifts with analytical tracking of the zenith offset) while validation is performed on canonical unaugmented volumes.
Within each fold, the binary classification threshold was determined strictly on the training split by maximizing Youden's index ($J = \text{TPR} - \text{FPR}$) without exposing validation samples. 
We report ROC AUC, Sensitivity, Specificity, F1-score, and the achieved validation Youden's index ($J_{\text{val}} = \text{Sensitivity} + \text{Specificity} - 1$), with 95\% confidence intervals (CI) computed via the Student's $t$-distribution across the 30 spatially perturbed evaluation runs ($df = 29$).
\section{RESULTS}

\subsection{Ablation of Backbone Adaptation Strategies in the Low-Data Regime}

To evaluate the risk of representation overfitting on the paired clinical cohort ($N = 212$ joints), we compared four parameter adaptation strategies for the BertsWin (PhysLoss + GCond) backbone on the synovial effusion detection task: (1) a strictly frozen encoder ($W_{\text{enc}} = \text{const}$) with only the spatial attention projector trained, (2) partial fine-tuning of the boundary layers (unfreezing the 3D CNN stem and final transformer block), (3) LoRA injected into the self-attention layers across ranks $r \in \{4, 8, 16\}$ ($\alpha = 16$) and $r = 32$ ($\alpha = 64$), and (4) full end-to-end fine-tuning with layer-wise learning rate decay. 

\begin{table}[ht]
    \centering
    \caption{Impact of backbone parameter adaptation strategies on cross-modal synovial effusion detection (10-fold CV).}
    \label{tab:adaptation_ablation}
    \small
    \setlength{\tabcolsep}{5pt}
    \begin{tabular}{lccc}
        \toprule
        \textbf{Adaptation Strategy} & \textbf{ROC AUC} & \textbf{Bal. Acc.} & \textbf{F1} \\
        \midrule
        \textbf{Frozen Encoder} & $\mathbf{0.793}$ & $\mathbf{0.730}$ & $\mathbf{0.593}$ \\
        {\footnotesize ($W_{\text{enc}} = \text{const}$)} & & & \\
        \addlinespace[3pt]
        Partial Fine-Tuning & $0.758$ & $0.712$ & $0.568$ \\
        {\footnotesize (Stem + Last Block)} & & & \\
        \addlinespace[3pt]
        LoRA Adaptation & $0.720$ & $0.685$ & $0.531$ \\
        {\footnotesize ($r \in \{4\dots32\}$)} & & & \\
        \bottomrule
    \end{tabular}
\end{table}

In preliminary runs, full end-to-end fine-tuning of all $86.28 \times 10^6$ parameters resulted in rapid memorization of patient-specific osseous anatomy within the initial training epochs, collapsing validation performance to the level of random guessing ($\text{ROC AUC} \approx 0.50$) and necessitating its exclusion from the final evaluation protocol. 
Among the parameter-restricted regimes (Table~\ref{tab:adaptation_ablation}), classification accuracy degraded monotonically as trainable degrees of freedom were introduced into the encoder. 
Even low-rank adapters, despite improving training-fold convergence on isolated splits, inflated inter-fold variance and reduced aggregate ROC AUC from $0.793$ to $0.720$. 
These results confirm that under extreme sample scarcity, granting the encoder any parametric flexibility shifts optimization toward high-amplitude individual bone contours rather than invariant intra-articular soft-tissue features, motivating strict weight freezing for all subsequent experiments. 

\subsection{Cross-Modal Detection of Synovial Effusion Across Pre-Trained Encoders}

We next investigated whether the sub-visual CBCT signature of MRI-verified synovial effusion can be extracted by standard self-supervised 3D encoders once isolated by our deterministic $8^3$ token cropping pipeline, or whether physics-guided pre-training is required to resolve the signal at $N = 212$ joints. 
Four frozen encoders were evaluated across identical patient-isolated $3 \times 10$-fold cross-validation splits: BertsWin (PhysLoss + GCond), BertsWin ($L_2$ + GCond), BertsWin (MVC + GCond), and the sparse MONAI MAE-ViT ($L_2$ + AdamW) baseline (Table~\ref{tab:effusion_results}). 

\begin{table*}[ht]
    \centering
    \caption{Cross-modal detection of MRI-verified synovial effusion across frozen pre-trained 3D encoders ($3 \times 10$ CV, mean $\pm$ 95\% CI).}
    \label{tab:effusion_results}
    \small
    \setlength{\tabcolsep}{5pt}
    \begin{tabular}{lcccccc}
        \toprule
        \textbf{Pre-Trained Encoder} & \textbf{ROC AUC} & \textbf{Sensitivity} & \textbf{Specificity} & \textbf{F1-Score} & 
        \begin{tabular}{@{}c@{}}\textbf{Youden's} \\ $\mathbf{J_{\text{val}}}$\end{tabular} & 
        \begin{tabular}{@{}c@{}}\textbf{Degenerate} \\ \textbf{Folds}\end{tabular} \\
        \midrule
        BertsWin (PhysLoss + GCond) & $\mathbf{0.793 \pm 0.036}$ & $\mathbf{0.697 \pm 0.052}$ & $\mathbf{0.762 \pm 0.062}$ & $\mathbf{0.593 \pm 0.039}$ & $\mathbf{0.459}$ & $\mathbf{2 / 30}$ \\
        BertsWin ($L_2$ + GCond) & $0.513 \pm 0.048$ & $0.442 \pm 0.118$ & $0.627 \pm 0.101$ & $0.297 \pm 0.077$ & $0.070$ & $7 / 30$ \\
        BertsWin (MVC + GCond) & $0.465 \pm 0.062$ & $0.394 \pm 0.115$ & $0.675 \pm 0.113$ & $0.281 \pm 0.079$ & $0.070$ & $8 / 30$ \\
        MAE-ViT ($L_2$ + AdamW) & $0.499 \pm 0.044$ & $0.603 \pm 0.104$ & $0.473 \pm 0.113$ & $0.365 \pm 0.050$ & $0.076$ & $5 / 30$ \\
        \bottomrule
    \end{tabular}
    \vspace{0.15cm}
    \parbox{\textwidth}{\footnotesize \textit{Note:} Metrics report mean $\pm$ 95\% CI aggregated across all 30 evaluation runs of the $3 \times 10$-fold cross-validation protocol ($df = 29$). While strict patient-level isolation between training and validation splits is enforced within every fold, the 30 runs do not represent 30 mutually independent patient cohorts, as the three 10-fold repeats re-evaluate the same 99 patients ($N = 212$ joints) across three different stratified patient partitions and training augmentation seeds. Under a conservative degree-of-freedom adjustment restricted to $K = 10$ non-overlapping patient blocks ($df = 9$, widening 95\% CIs by a factor of $1.92$), the lower 95\% CI bound of ROC AUC for BertsWin (PhysLoss + GCond) is $0.724$, remaining decisively above the $0.50$ random-chance threshold. \textit{Degenerate Folds} denotes validation splits that failed convergence sanity checks (single-class threshold collapse with $\text{threshold} = \infty$, or numerical divergence) due to small-sample noise. To avoid survivor bias and preserve the full 30-run sample size ($df = 29$), failed splits were not discarded; instead, they were retained in the aggregation using a conservative fallback state ($\text{Sensitivity} = 0.0$, $\text{Specificity} = 1.0$, $\text{F1} = 0.0$, and the achieved $\text{AUC}$, or $\text{AUC} = 0.50$ upon numerical failure).}
\end{table*}

Because a cohort of 212 joints ($\sim 21$ validation joints per fold) operates near the lower statistical bound required for 3D cross-modal learning, individual fold stability serves as a direct indicator of the underlying feature signal-to-noise ratio: 
\begin{itemize}
    \item \textbf{Sample-Scarcity Vulnerability of Unmasked Objectives:} While standard $L_2$ or global MVC encoders might potentially resolve soft-tissue attenuation patterns if trained on thousands of paired volumes, their latent spaces did not provide a sufficient signal-to-noise ratio to prevent classifier overfitting at $N = 212$. 
    Consequently, BertsWin ($L_2$ + GCond), BertsWin (MVC + GCond), and MONAI MAE-ViT ($L_2$ + AdamW) yielded aggregate ROC AUC scores between $0.465$ and $0.513$, with validation Youden's indices ($J_{\text{val}} = 0.070\text{--}0.076$) statistically indistinguishable from random guessing. 
    Inspection of individual splits revealed that in $7/30$ folds for BertsWin ($L_2$) and $8/30$ folds for BertsWin (MVC), noise overfitting inverted the training ROC curve ($\text{TPR} < \text{FPR}$), causing Youden threshold selection to default to a degenerate boundary ($\text{threshold} \to \infty$, $\text{Sensitivity} = 0.0$, $\text{F1} = 0.0$), whereas MONAI MAE-ViT experienced numerical convergence failure on $3/30$ folds and threshold collapse on an additional $2/30$ folds. 
    \item \textbf{Signal Recovery via PhysLoss + GCond:} Pre-training with the anatomy-aware PhysLoss objective and GCond did not completely immunize the classifier against small-sample variance---$2/30$ folds still exhibited localized noise overfitting due to the restricted validation block size. 
    Nevertheless, BertsWin (PhysLoss + GCond) was the \textit{only} configuration to overcome this sample-size bottleneck across the 30-fold ensemble, achieving a statistically robust $\text{ROC AUC} = 0.793 \pm 0.036$ (conservative $df=9$ lower 95\% CI bound of $0.724$), $\text{Sensitivity} = 0.697 \pm 0.052$, $\text{Specificity} = 0.762 \pm 0.062$, and Youden's $J_{\text{val}} = 0.459$. 
\end{itemize}

\begin{figure}[h]
    \centering
    \includegraphics[width=\columnwidth]{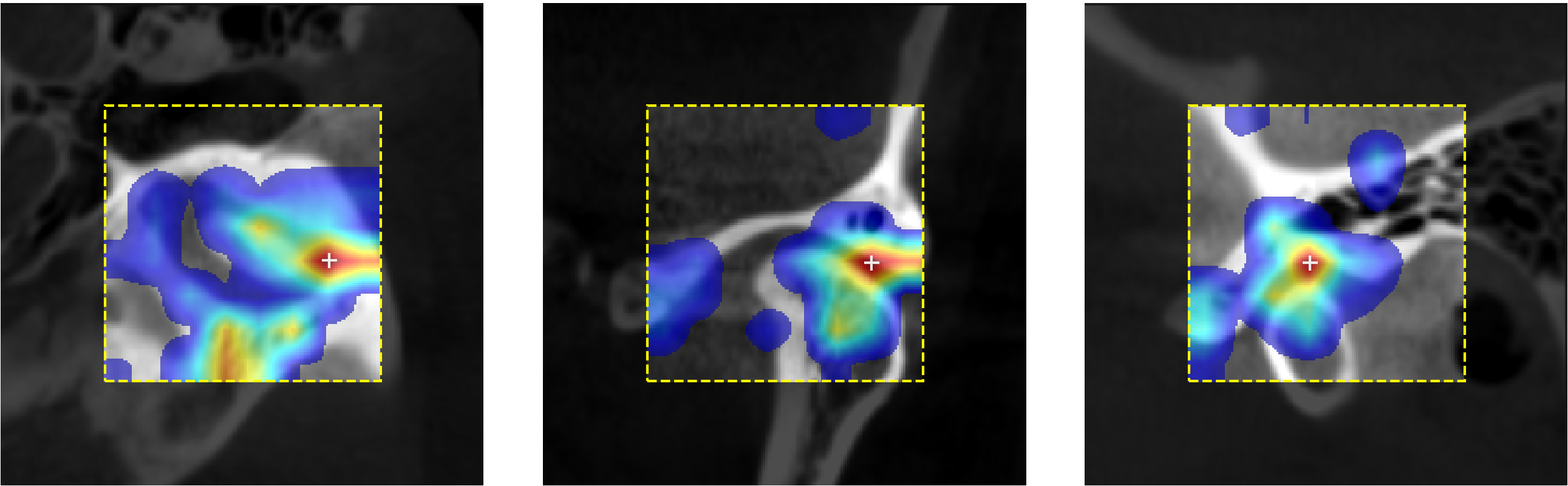}
    \caption{Grad-CAM attention maps for cross-modal synovial effusion detection across axial, coronal, and sagittal CBCT views. The yellow dashed bounding box delineates the cropped $8^3$ token ROI; the white crosshair marks peak network activation localized strictly within the intra-articular synovial cavity.}
    \label{fig:gradcam_effusion}
\end{figure} 

To verify that the BertsWin (PhysLoss + GCond) classifier relied on intra-articular physical features rather than spurious background correlations, Gradient-weighted Class Activation Mapping (Grad-CAM) was performed across the validation cohort (Fig.~\ref{fig:gradcam_effusion}). 
Across all three orthogonal planes, peak gradient activation localized strictly within the radiolucent joint space between the condylar head and glenoid fossa, matching the anatomical distribution of synovial fluid effusion.

\subsection{Articular Disc Displacement Detection via Osseous Gradient Priors}

In contrast to synovial effusion, direct classification of articular disc displacement solely from frozen soft-tissue latent features failed to reach statistical significance on the 212-joint cohort, as the minimal attenuation differential between the thin fibrocartilaginous disc and adjacent musculature fell below the noise floor at this sample scale. 
Integrating the analytical vertical spatial gradient ($|\nabla_z I|$) filter alongside the cropped BertsWin (PhysLoss + GCond) features stabilized convergence and yielded statistically significant discrimination (Table~\ref{tab:disc_results}). 

\begin{table}[ht]
    \centering
    \caption{Comparison of direct soft-tissue effusion detection and gradient-guided articular disc displacement detection using BertsWin (PhysLoss + GCond) ($3 \times 10$ CV, mean $\pm$ 95\% CI).}
    \label{tab:disc_results}
    \small
    \setlength{\tabcolsep}{4pt}
    \begin{tabular*}{\columnwidth}{@{\extracolsep{\fill}}lcc@{}}
        \toprule
        \textbf{Metric} & \textbf{Synovial Effusion} & \textbf{Disc Displacement} \\
        \midrule
        ROC AUC     & $0.793 \pm 0.036$ & $0.679 \pm 0.049$ \\
        Sensitivity & $0.697 \pm 0.052$ & $0.840 \pm 0.037$ \\
        Specificity & $0.762 \pm 0.062$ & $0.585 \pm 0.078$ \\
        F1-Score    & $0.593 \pm 0.039$ & $0.838 \pm 0.022$ \\
        \bottomrule
    \end{tabular*}
\end{table}

\begin{figure}[ht]
    \centering
    \includegraphics[width=\columnwidth]{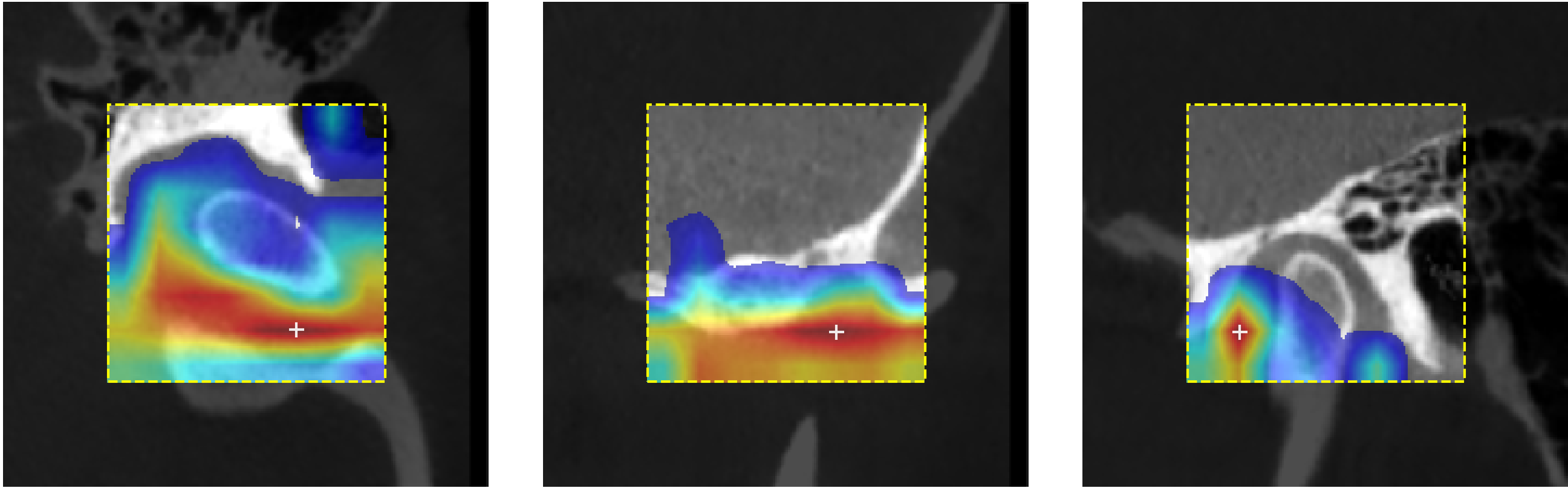}
    \caption{Grad-CAM attention maps for articular disc displacement detection with the vertical spatial gradient ($|\nabla_z I|$) prior across axial, coronal, and sagittal CBCT projections. Peak activation shifts from the joint cavity to the cortical bone surfaces of the condyle and articular eminence, reflecting secondary osseous remodeling.}
    \label{fig:gradcam_disc}
\end{figure}

With the vertical spatial gradient prior, the classifier achieved $\text{ROC AUC} = 0.679 \pm 0.049$ (conservative $df=9$ CI of $\pm 0.094$, validation Youden's $J_{\text{val}} = 0.425$), characterized by a pronounced asymmetry between high sensitivity ($\text{Sensitivity} = 0.840 \pm 0.037$, $\text{F1} = 0.838 \pm 0.022$) and moderate specificity ($0.585 \pm 0.078$). 
Grad-CAM inspection (Fig.~\ref{fig:gradcam_disc}) confirmed that the vertical spatial gradient ($|\nabla_z I|$) shifted network attention away from the radiolucent joint space onto the high-contrast cortical contours of the condylar head and articular eminence. 
Consequently, the model detected chronic secondary osteoarthritic remodeling and condylar flattening associated with disc derangement, explaining both its high sensitivity for displaced discs and its elevated false-positive rate in patients whose bone contours remained permanently remodeled despite prior clinical disc reduction. 

\section{DISCUSSION}

\subsection{Overcoming the Small-Data Bottleneck in Cross-Modal Clinical Research}

A critical nuance in interpreting our experimental results (Table~\ref{tab:effusion_results}) is understanding why encoders pre-trained with standard $L_2$ loss or unmasked multi-component variance (MVC) failed to exceed random guessing ($\text{ROC AUC} = 0.465\text{--}0.513$) on the synovial effusion task. 
This outcome should not be misconstrued as evidence that $L_2$- or MVC-based representations are inherently incapable of encoding soft-tissue features. 
Given a massive paired training set of several thousand synchronous CBCT--MRI volumes, even a standard sparse MAE-ViT backbone might eventually isolate weak cross-modal attenuation correlations. 
Rather, the collapse of unmasked baselines---as well as the localized overfitting observed on $2/30$ validation folds even for our best-performing PhysLoss configuration---is a direct consequence of operating at the extreme lower statistical bound of sample size ($N = 212$ joints, or $\sim 21$ joints per validation fold). 

Empirically, optimization in this low-SNR regime exhibits a sharp early-epoch bifurcation: if the projection head fails to lock onto the weak intra-articular signal during the initial training epochs, it never recovers, converging irreversibly toward stochastic noise and patient-specific osseous anatomy. In fact, even within an identical fold partition, simply permuting the queue order in which patients are sampled into early mini-batches can determine whether the trajectory enters the diagnostic basin of attraction or collapses into a degenerate local minimum.

However, in real-world clinical research, assembling thousands of paired multimodal scans under strict temporal synchronization ($\le 30$~days between CBCT and MRI to prevent natural disease progression from corrupting ground-truth labels) is logistically and financially prohibitive for independent academic groups and specialized clinical centers. 
Consequently, developing sample-efficient paradigms that can reliably extract weak diagnostic signals from cohorts of a few hundred patients is a practical necessity. 
Our ablation experiments demonstrate that succeeding in this micro-data regime requires eliminating superfluous variance at every stage of the computational pipeline: freezing the $86.28 \times 10^6$-parameter backbone to prevent memorization of individual osseous contours (Table~\ref{tab:adaptation_ablation}) and applying deterministic 3D token cropping ($14^3 \to 8^3$) around the condylar zenith to discard 81.3\% of background anatomical noise. 

\subsection{Physical Calibration of the Hounsfield Scale and Spectral Mechanics of BertsWin}

Why did combining \textbf{BertsWin}, \textbf{PhysLoss}, and \textbf{GCond} succeed in extracting the sub-visual effusion signature at $N = 212$ where all other pre-training regimes failed? 
The answer lies in the synergy between the intrinsic spectral behavior of the full-grid BertsWin architecture and the physical calibration of X-ray computed tomography. 

In standard representation learning, a higher effective covariance rank ($R_{\text{eff}}$) is conventionally interpreted as a marker of feature richness. 
However, our spectral audits across both maxillofacial CBCT and brain MRI benchmarks \cite{LimarenkoBertsWin} reveal a critical pathology in sparse 3D masked autoencoders (MONAI MAE-ViT): when 75\% of tokens are discarded, the encoder lacks a continuous 3D spatial framework and expends its representational dimensionality on memorizing high-frequency stochastic noise and spurious background activations. Remarkably, on skull-stripped brain MRI---where the background is homogeneous zero-valued space---sparse MAE-ViT models still generate an inflated effective rank across the whole volume that drops by more than half when evaluated strictly inside the brain ROI (e.g., $41.74 \to 21.44$), accompanied by severe representation anisotropy ($\sim 10^{-8}$). 
Similarly, on TMJ CBCT data, MAE-ViT ($L_2$) yields a high global effective rank ($R_{\text{eff}} = 23.36$) but fails to capture true anatomical orientation or intra-patient consistency ($0.329$). 

In stark contrast, maintaining a complete $14^3$ token grid in BertsWin acts as a structural inductive bias that suppresses background noise memorization while preserving latent space isotropy orders of magnitude higher than sparse ViTs \cite{LimarenkoBertsWin}. 
When coupled with variance-decomposed objectives (MVC) and the GCond optimizer on soft-tissue-dominant modalities such as brain MRI, BertsWin keeps global background rank compact while selectively expanding the effective dimensionality \textit{inside} the anatomical ROI to encode fine structural gradients. 
Yet on CBCT, where mineralized bone spans nearly $3000$~HU and dominates global variance, unmasked MVC optimization still gets hijacked by high-contrast cortical boundaries, leaving soft-tissue features under-resolved ($\text{ROC AUC} = 0.465$). 

Crucially, unlike MRI intensities, CT and CBCT attenuation values are anchored to the physically calibrated Hounsfield scale, where water, synovial fluid, and non-mineralized soft tissues consistently reside within a known density window ($[-300, 300]$~HU). 
By leveraging this physical prior, \textbf{PhysLoss} isolates the MVC structural correlation term inside the $M_{\text{soft}}$ mask ($\lambda_{\text{soft}} = 0.5$), while GCond's sign-based layer-wise scaling ensures that weak gradients from this low-contrast regime propagate across all 12 transformer blocks without vanishing. 
This forces the encoder to filter out high-frequency CBCT photon scatter (compressing global rank to a noise-free core of $R_{\text{eff}} = 4.01$) while preserving true 3D morphological chirality ($p < 0.001$) and intra-articular density variations. 
Consequently, when cropped to the $8^3$ peri-articular sub-volume, the frozen PhysLoss + GCond features provide a sufficiently clean signal-to-noise ratio for the attention projector to detect the subtle X-ray attenuation shift produced by synovial effusion ($\text{ROC AUC} = 0.793$, Fig.~\ref{fig:gradcam_effusion}). 

\subsection{Direct Fluid Sensing vs.\ Secondary Osseous Remodeling}

The divergence in classification behavior between synovial effusion and articular disc displacement (Table~\ref{tab:disc_results}) illustrates the physical limits of cross-modal CBCT inference at our current cohort scale. 
Synovial effusion represents a volumetric accumulation of inflammatory exudate throughout the joint cavity, altering bulk X-ray attenuation across multiple contiguous voxels within the $[-300, 300]$~HU band. 
Because PhysLoss explicitly sensitizes the encoder to structural correlations in this exact density range, the classifier detects effusion directly within the radiolucent joint space (Fig.~\ref{fig:gradcam_effusion}). 

Conversely, the articular disc is a thin fibrocartilaginous plate whose baseline attenuation is nearly indistinguishable from the adjacent capsule and muscle insertions. 
Although the $M_{\text{surf}}$ shell ($\lambda_{\text{surf}} = 0.2$) in PhysLoss was designed to capture the osseous-to-soft-tissue transition zone housing the disc, the direct radiographic contrast of fibrocartilage proved too weak to resolve on a sample of 212 joints without overfitting. 
Introducing the analytical vertical spatial gradient ($|\nabla_z I|$) prior successfully recovered statistically significant discrimination ($\text{ROC AUC} = 0.679 \pm 0.049$, $\text{Sensitivity} = 0.840$), but fundamentally altered the diagnostic mechanism: as confirmed by Grad-CAM (Fig.~\ref{fig:gradcam_disc}), the classifier shifted its focus to secondary osteoarthritic remodeling and flattening of the cortical bone surfaces \cite{Gorucu2019, Osama2025}. 
Because chronic osseous remodeling persists permanently on CBCT even after a displaced disc has been clinically reduced to a normal position on MRI, reliance on secondary bony markers yields high sensitivity ($0.840$) but bounds specificity ($0.585$) due to false-positive detections in historically remodeled joints. 
We posit that direct soft-tissue detection of disc displacement on CBCT remains theoretically feasible with PhysLoss representations, but will require a substantially larger paired training cohort to separate fibrocartilage attenuation from background quantum noise. 

\subsection{Broader Implications for Opportunistic Screening and Limitations}

Beyond maxillofacial radiology, our findings establish a broader methodological blueprint for \textbf{opportunistic screening of occult soft-tissue pathologies} during routine X-ray tomography. 
In daily clinical practice, millions of low-contrast CT and CBCT scans are acquired for routine dental, orthopedic, or trauma indications. 
While dangerous non-calcified abnormalities---such as early soft-tissue neoplasms, perineural infiltrations, or localized deep-space inflammations---are physically captured within the reconstructed field of view, their subtle attenuation signatures remain invisible to the human eye under standard bone or wide-window viewing protocols. 
Our work demonstrates that by combining a topology-preserving 3D self-supervised backbone (BertsWin), a physics-guided loss restricted to the target soft-tissue HU window (PhysLoss), and automated anatomical cropping around segmented landmarks, routine diagnostic scans can be repurposed as background screening tools to flag sub-visual soft-tissue anomalies for targeted MRI follow-up. 

Several limitations of the present study must be noted: 
\begin{enumerate}
    \item \textbf{Sample Size Constraints:} While our cohort of 212 paired joints (99 patients) is strictly curated via a $\le 30$-day inter-scan window and heterogeneous multi-scanner inclusion (6 devices), it lies at the minimal statistical threshold for 3D cross-validation, precluding the evaluation of rare pathologies with severe class imbalance (such as cortical destruction at $1:4$). 
    \item \textbf{Anatomical Field-of-View Restriction:} Cropping the latent grid from $14^3$ to $8^3$ tokens around the condylar zenith was essential to boost the intra-articular signal-to-noise ratio at $N = 212$, but inevitably excludes extra-articular soft tissues---most notably the lateral pterygoid muscle, whose functional spasm and structural changes are closely linked to TMJ internal derangement. 
    \item \textbf{Residual Inter-Scan Dynamics:} Although a 30-day maximum interval between CBCT and MRI is three times stricter than routine clinical windows, acute physiological fluctuations in synovial effusion volume between scans cannot be completely excluded and may contribute to residual label noise. 
\end{enumerate}

\section{CONCLUSION}

This study demonstrates that visually imperceptible, MRI-verified soft-tissue pathologies of the temporomandibular joint can be detected directly from routine 3D CBCT scans under extreme paired-data scarcity ($N = 212$ joints). 
We show that overcoming noise memorization in this low-data regime requires the strict integration of three principles: (1) self-supervised pre-training of a full-grid 3D encoder (BertsWin) with the GCond optimizer and an anatomy-aware objective (PhysLoss) that exploits the calibrated Hounsfield scale ($[-300, 300]$~HU) to isolate soft-tissue structural correlations while filtering out spectral scatter; (2) freezing the pre-trained backbone ($86.28 \times 10^6$ parameters) to prevent overfitting to individual osseous geometry; and (3) deterministic 3D token cropping ($14^3 \to 8^3$) anchored on the automatically segmented condylar zenith to eliminate 81\% of background variance. 
Whereas standard $L_2$ and unmasked MVC encoders fail to extract a statistically valid signal ($\text{ROC AUC} = 0.465\text{--}0.513$), our physics-guided pipeline achieves $\text{ROC AUC} = 0.793 \pm 0.036$ for direct synovial effusion detection and $\text{ROC AUC} = 0.679 \pm 0.049$ ($\text{Sensitivity} = 0.840$) for articular disc displacement via secondary cortical gradient priors. 
These results establish a generalizable framework for extracting sub-visual soft-tissue biomarkers from routine radiographic procedures, paving the way for opportunistic cross-modal screening in clinical radiology. 

\section{DATA AVAILABILITY}
\begin{itemize}
    \item BertsWinMAE: \url{https://github.com/AlevLab-dev/BertsWinMAE} 
    \item GCond: \url{https://github.com/AlevLab-dev/GCond} 
\end{itemize}

\section*{ACKNOWLEDGMENTS}
During the preparation of this manuscript, the author used Google Gemini for Russian-to-English translation and language editing. The author reviewed, verified, and edited all translated text and takes full responsibility for the content of the publication.

\end{document}